\documentclass[twocolumn,prc,aps,nofootinbib,showkeys,superscriptaddress]{revtex4}

\usepackage[latin9]{inputenc}
\usepackage{color}
\usepackage{amsmath}
\usepackage{amsthm}
\usepackage{amssymb}
\usepackage{graphicx}
\usepackage{hyperref}
\makeatletter
\usepackage{epsfig}
\usepackage{amsthm}
\usepackage{tikz}
\usetikzlibrary{quantikz}
\usepackage{comment}
\usepackage{array}
\usepackage{booktabs}
\usepackage{ulem}
\usepackage{diagbox}
\usepackage{multirow}

\usepackage{bm}

\usepackage{color}

\makeatother

\begin{document}

\title{Neutron-proton pairing correlations described on quantum computers}

\author{Jing Zhang}
\email{jing.zhang@ijclab.in2p3.fr}
\affiliation{Universit\'e Paris-Saclay, CNRS/IN2P3, IJCLab, 91405 Orsay, France}

\author{Denis Lacroix}
\email{denis.lacroix@ijclab.in2p3.fr}
\affiliation{Universit\'e Paris-Saclay, CNRS/IN2P3, IJCLab, 91405 Orsay, France}

\author{Yann Beaujeault-Taudi\`ere\footnote{Current affiliation:  Alice \& Bob, 49 boulevard du G\'en\'eral Martial Valin, 75 015 Paris, France}}
\affiliation{Universit\'e Paris-Saclay, CNRS/IN2P3, IJCLab, 91405 Orsay, France}
\affiliation{Laboratoire Leprince-Ringuet (LLR), \'Ecole polytechnique, CNRS/
 IN2P3, F-91128 Palaiseau, France}

\date{\today}

\begin{abstract}
The ADAPT-VQE approach is used to solve the neutron-proton pairing problem in atomic nuclei. This variational approach is considered today as one of the most powerful methods to iteratively find the ground state of a many-body problem, provided a performing set of operators, called the pool of operators, is used to explore the Hilbert space of many-body wave-functions. Three different pools of operators, which might eventually break one or several symmetries of the Hamiltonian during the descent to the ground state, are tested for the neutron-proton pairing problem. We observe that the breaking of some symmetries during the optimization of the trial wave-function might, in general, help to speed up the convergence towards  
the ground state. Still, we rejected the pool of operators that might explicitly break the total particle number because they become uncontrollable during the optimization process. Overall, we observed that the iterative optimization process 
rapidly becomes a delicate problem when the number of parameters to build the ansatz increases, and the energy might get stuck at energies higher than the ground state energy. To improve the convergence in this case, several techniques have been proposed, with some better controlling the symmetries during the energy minimization. Among the proposed methods, two have proven effective: one based on an embedding technique and the other on a randomized preparation of the initial state. We conclude that the ADAPT-VQE, complemented by these techniques, can provide a very accurate description of the neutron-proton pairing problem, 
and can outperform other standardly used techniques that break the particle number symmetry and restore afterwards.           
\end{abstract}

\maketitle 

\section{Introduction}

Following the recent trend in various fields of physics and chemistry \cite{McC16,Fan19,Cao19,McA20,Bau20,Ayr23,Dim24}, active
research is being conducted to use quantum computers for low-energy nuclear structure studies \cite{Dum18,Lu19,Rog19,Ste22,Rom22,Tur22}. 
Provided that quantum computers can reach sufficient accuracy, this novel kind of processor can significantly surpass
classical computers for nuclear structure calculations. Quantum algorithms such as Quantum Phase estimation (QPE) \cite{Nie00} potentially 
allow solving eigenvalue problems in huge Hilbert spaces and might, in the future, open the possibility of performing exact ab initio calculations 
on the whole chart of atomic nuclei. Unfortunately, due to the current status of quantum technologies, such perfect solvers cannot yet be employed. 

Most studies today use variational techniques like the Variational Quantum Eigensolver (VQE) \cite{Per14,Bha21,End21,Ayr23}, 
which are more robust to noisy devices. In nuclear physics, the majority of the current studies are exploring the different techniques for accurately 
preparing the ground state (GS) of atomic nuclei using specific quantum ansatzes \cite{Dim21,Kis22,Sar23,Per23,Rob23,Yos24,Bho24,Bai24,Gib24}. 
Various techniques have also been introduced to access excited states \cite{Hla22,Gro23,Hla24,Bea24}. 
One important question to be clarified in this context is the necessity of using a symmetry breaking/symmetry restoration (SB/SR) technique 
to design a nuclear physics ansatz. Such a strategy is commonly used in the nuclear density functional context to prepare accurate ansatz
at low numerical cost \cite{Rin80,Bla86}. A typical example is the pairing problem, where a BCS ansatz 
breaking the $U(1)$ symmetry associated with the particle number conservation \cite{Rin80,Bla86,Bri05} might be used. This direction has already 
been explored in the context of quantum computing for the pairing problem \cite{Lac20,Tsu20,Siw21,Kha21,Rui22,Tsu22,Rui23,Rui24}, including the 
possibility of restoring the broken symmetries \cite{Lac23}. However, previous studies were restricted to particle-like pairing. 

Here, we address the more general problem of proton-neutron pairing in atomic nuclei, where both isospin and spin channels are treated. SB/SR strategies have already been 
widely investigated for proton-neutron pairing in the context of classical computing 
\cite{Goo79,Fra14}. At the early stage of the present work, we also extensively explored this direction for quantum computations, allowing the possibility to form different spin-isospin pairs in the ansatz \cite{JingPhD}. 
The first conclusion of the present work, which will not be further developed in the core of the article, is that SB/SR strategy might lead to significant difficulties when too many 
symmetries are simultaneously broken. For the specific neutron-proton problem, we encountered the following issues: (i) First, the generalized BCS ansatz introduced 
for spin-isospin pairing turns out to be difficult to converge in practice using standard optimizers in a quantum hybrid classical-quantum calculation. (ii) Because of the enlarged single-particle space in the proton-neutron pairing case, our tests were restricted 
to rather small numbers of particles. Then, the converged SB state presented significant fluctuations in particle number compared 
to the mean number of particles, which is unsatisfactory; and (iii) we also implemented the symmetry-restoration
techniques proposed in Refs. \cite{Lac20,Siw21,Rui22,Lac23}. When the SB state varies, both the total spin and total isospin, as well as neutron and proton numbers might simultaneously be broken, requiring to restore all these four symmetries. The simultaneous restoration of these symmetries, even after the convergence (Projection After Variation), while possible, turns out to be extremely costly on quantum computers. 

Intending to prepare future applications in nuclear physics, we focus the present study on the use of a variational technique and, more specifically, the promising Adaptive Derivative-Assembled Pseudo-Trotter ansatz-Variational Quantum Eigensolver (ADAPT-VQE) approach \cite{Gri19}, to treat the neutron-proton pairing problem. This problem is a perfect playground for testing variational ansatzes since it already contains many competing interaction channels and symmetries, making determining the ground state particularly challenging. It is also an important intermediate milestone for future complete
shell-model calculations, in that it allows the test of the expressive power and 
convergence patterns for  
different variational ansatzes. Specifically, we identify cases 
where convergence might be difficult and propose 
a set of methods to overcome some of the difficulties encountered 
when applying the ADAPT-VQE technique to the case of several competing interaction channels coexisting in Hamiltonians. 

In section \ref{sec:nppairing}, we briefly recall the neutron-proton pairing Hamiltonian. In section \ref{sec:adaptvqe}, the ADAPT-VQE and several pools of operators leading to different 
types of ansatzes are introduced. The method is also illustrated for the particle-like pairing case and compared with the SB/SR strategy. In section \ref{sec:adaptappliednp}, the ADAPT-VQE technique is applied to a variety of neutron-proton problems, in which we switch on and off different spin-isospin channels to assess the predictive power of the approach and identify potential difficulties. In section \ref{sec:improve}, we introduce and discuss several new methods to overcome some problems encountered in describing neutron-proton pairing correlations on a quantum processor.

\section{neutron-proton pairing Hamiltonian}
\label{sec:nppairing}

Here, we consider an ensemble of neutrons and protons that can access a set of single-particle states. 
We assume 
time-reversal symmetry and denote by $\bar i$ the time-reversed state of $i$, both having the same 
energy. The particles are interacting through the spin-isospin pairing Hamiltonian:
\begin{eqnarray}
H&=& \sum_{i=1}^{n_B}  \left[\varepsilon_{i, n} (\nu^\dagger_i \nu_i + \nu^\dagger_{\bar i} \nu_{\bar i}) + 
\varepsilon_{i, p}(\pi^\dagger_i \pi_i + \pi^\dagger_{\bar i} \pi_{\bar i}) \right]
\nonumber \\
&-& 
\sum_{T_z = -1,0,1}  g_V(T_z) {\cal P}^\dagger_{T_z} {\cal P}_{T_z} \nonumber \\
&-& 
\sum_{S_z = -1,0,1} g_S(S_z) {\cal D}^\dagger_{S_z} {\cal D}_{S_z}.   \label{eq:hamilNP}
\end{eqnarray}
In these expressions, the creation operators $\nu^\dag_{i/ \bar i}$ (resp. $\pi^\dag_{i/ \bar i}$) are associated 
with the neutron (resp. the proton) single-particle states. 
$n_B$ denotes the number of blocks, where each block consists of 4 single-particle states for different spin-isospin components.   
The second line denotes the $(S,T) = (0,1)$ isosvector pairing channels and the third line depicts the isoscalar ones of $(S,T) = (1,0)$, where $S$ and $T$ are the total spin and isospin quantum number of a particle pair. $S_z$ and $T_z$ designate the spin and isospin projection. Finally, $g_S(S_z)$ and $g_V(T_z)$ are six two-body coupling constants describing the strength of different pairing interaction channels.
The pair creation operator of a given channel decomposes as that of individual pairs formed on specific levels in each block:
\begin{eqnarray}
 {\cal P}^\dagger_{T_z} &=& \sum_{i} P^\dagger_{T_z, i}, ~~ {\cal D}^\dagger_{S_z} = \sum_{i} D^\dagger_{S_z, i}.
\end{eqnarray}
 Specifically, we have:
\begin{eqnarray}
\left\{ 
\begin{array}{l}
\displaystyle P^\dagger_{1,i} =  \nu^\dagger_{ i} \nu^\dagger_{\bar i} , ~~~~
\displaystyle P^\dagger_{-1,i} = \pi^\dagger_{i} \pi^\dagger_{\bar i} \\
\\
\displaystyle P^\dagger_{0,i} = \frac{1}{\sqrt{2}}  \left[ \nu^\dagger_{i} \pi^\dagger_{\bar i}  +\pi^\dagger_{i} \nu^\dagger_{\bar i}   \right]  \\
\\
\displaystyle D^\dagger_{1,i} =  \nu^\dagger_{i} \pi^\dagger_{ i}, ~~~~
\displaystyle  D^\dagger_{-1,i} =\nu^\dagger_{\bar i} \pi^\dagger_{\bar i} \\
\\
\displaystyle D^\dagger_{0, i} = \frac{1}{\sqrt{2}} \left[ \nu^\dagger_{i} \pi^\dagger_{\bar i}  -\pi^\dagger_{i} \nu^\dagger_{\bar i}   \right] \\
\end{array}
\right. \label{eq:pairoperators}
,
\end{eqnarray}
where, by convention, we simply assume that $(i, \bar i)$ are associated with the single-particle spins $s_z = (\uparrow, \downarrow)$. It is also assumed that the convention for isospin component $\tau_z$ is $\frac{1}{2}$ (resp. $-\frac{1}{2}$) for neutrons (resp. for protons).

\begin{table}[htbp]
\newcolumntype{C}[1]{>{\centering\arraybackslash}p{#1}} 
\newlength{\colwidth}
\setlength{\colwidth}{.7cm}
\begin{tabular}{|C{2cm}@{\hspace{.0cm}}|C{\colwidth}C{\colwidth}C{\colwidth}@{\hspace{.3cm}}C{\colwidth}C{\colwidth}C{\colwidth}|}
   \hline
  \multirow{2}{*}{\backslashbox{Case}{$S_z/T_z$}}   & \multicolumn{3}{c}{Isoscalar}   & \multicolumn{3}{c|}{Isovector} \\
      & $-1$ & 0 & 1 &  $-1$ & 0 & 1 \\
    \hline
    1 &            &            &            & \checkmark &            & \checkmark \\
    2 &            & \checkmark &            &            & \checkmark & \\
    3 &            &            &            & \checkmark & \checkmark & \checkmark \\
    4 & \checkmark & \checkmark & \checkmark & \checkmark & \checkmark & \checkmark \\
    \hline
  \end{tabular}  
  \caption{Schematic table presenting the terms that are retained (terms with ``\checkmark'' in the table) for the different variants of the Hamiltonian used in this work. The different cases are labelled from 1 to 4 in the main text, and correspond to different cases of interest in nuclear physics. 
  {\bf Case~1}: The particle-like pairing only Hamiltonian. This case consists in only the ${\cal P}^\dagger_1$ and ${\cal P}^\dagger_{-1}$ terms, i.e. only neutron-neutron and proton-proton interactions. It is the typical choice for nuclear systems when using energy density functional theory. 
  An illustration of such calculation is given in Fig. \ref{fig:pairing-nn}.  
  \textbf{Case~2}: The case retaining only the $T_z=S_z=0$ components. 
  \textbf{Case~3}: The case where only the channel of $(S,T) = (0,1)$ is considered. This case corresponds to $g_S=0$ in Eq. (\ref{eq:hamilNP}). 
  \textbf{Case~4}: Full proton-neutron pairing Hamiltonian. 
  Case 1 to 4 will sometimes be referred to as 1: ($T_z = \pm 1$); 2: ($S_z=T_z=0$); 3: ($S=0$) only; and 4: (Full) in the text.
}
  \label{tab:Hvariants}
\end{table}

Our objective is to systematically investigate some of the cases commonly encountered 
in nuclear physics. More specifically, the different cases correspond to different choices of the two-body coupling constants in the general Hamiltonian given by Eq.~(\ref{eq:hamilNP}).
We consider four situations that are representative of atomic nuclei, and list them in table \ref{tab:Hvariants}. The most common one, especially in nuclear density functional theory, is the ``case 1'' in the table,
where only like particles can form pairs, i.e. neutron-neutron or proton-proton pairs. In the subsequent part of the article, the couplings retained in each case, as listed in table \ref{tab:Hvariants}, are consistently assigned a uniform coupling strength $g_S(S_z) = g_V(T_z) =g$ for all $S_z$ and $T_z$ values, unless specified differently. 

\section{Discussion of the ADAPT-VQE approach}
\label{sec:adaptvqe}

As announced in the introduction, we mainly focused on the ADAPT-VQE approach \cite{Gri19}. This approach is versatile and is considered today as one of the most powerful techniques for near-term 
applications on noisy devices. Its properties and possible difficulties, as well as associated solutions, have also been widely discussed in the literature \cite{Tan21,Yao21,Yor21,Zha21,Sma21,Hai22,Liu22,Yor22,Ber23,Fen23,Gri23,Ana24}.  
After briefly describing the technique, we give a first illustration of its power on the pairing problem.  

\subsection{Description of the ADAPT-VQE approach}
\label{sec:adaptgen}

We explore below the ADAPT-VQE technique to approximate the ground state of
the general proton-neutron Hamiltonian given by Eq.~(\ref{eq:hamilNP}). This technique was employed in recent years, for instance, in the Lipkin \cite{Rom22} and Agassi models \cite{Bai24} with some successes in reproducing permutation invariant systems. This method is already well 
documented, and we only give the main steps here. Starting from an initial state $| \Psi_0 \rangle$ and a ``pool'' of pre-selected operators $\{ G_\alpha\}_{\alpha = 1, \Omega}$ with $\Omega$ the pool size, the method builds iteratively a trial wave-function of the form 
\begin{eqnarray}
| \Psi_n \rangle &=& \prod_{i=1}^{n_{\rm step}} e^{i \theta_i G_{\alpha_i}} | \Psi_0 \rangle .\label{eq:trialadapt}
\end{eqnarray} 
In its simplest implementation, the ADAPT-VQE approach proceeds as follows:
\begin{enumerate}
    \item We chose an initial state $| \Psi_0 \rangle$ that can be conveniently encoded on a digital quantum computer.
    \item Starting from this state, we seek for a new state 
    $| \Psi_1 \rangle= e^{i \theta_1 G_{\alpha_1}}| \Psi_0 \rangle$, where $G_{\alpha_1}$ is chosen within the operator 
    pool such that the variation of the energy $E_1 \equiv \langle \Psi_1 | H | \Psi_1 \rangle$ with respect to $\theta_1$,  
    \begin{eqnarray}
\left.\frac{\partial E_1}{\partial \theta_1}\right|_{\theta_1=0} &=&i  \langle \Psi_0 |\left[ H, G_{\alpha_1}\right]| \Psi_0 \rangle,   \label{eq:grad}
\end{eqnarray}
    is maximized. Once the operator $G_{\alpha_1}$ is identified, the associated ansatz is optimized by minimizing the energy through the standard VQE method with respect to the parameter 
    $\theta_1$. 
    \item The process is then iterated. Assume that at a given step $n-1$, a sequence of operators $\{ G_{\alpha_1}, \cdots , G_{\alpha_{n-1}}\} $ has been selected, and the set of variational parameters $\{ \theta_1 , \cdots , \theta_{n-1}\}$
    has been optimized to minimize the energy $E_{n-1}$. Then, for the next step, the new trial state is given by $| \Psi_n \rangle = e^{i \theta_n G_{\alpha_{n}}}| \Psi_{n-1} \rangle$, where $G_{\alpha_n}$ is the operator among the operators pool that maximizes the energy gradient:
    \begin{eqnarray}
\left. \frac{\partial E_n }{\partial \theta_n}\right|_{\theta_n=0} &=&i  \langle \Psi_{n-1} |\left[ H, G_{\alpha_n} \right]| \Psi_{n-1} \rangle. \label{eq:gradient}
\end{eqnarray}
After adding this new operator to the sequence of previously selected operators, the entire set of parameters $\{ \theta_1, \cdots , \theta_n\}$ is varied to minimize the energy through the VQE procedure. As will be discussed below, re-optimizing the full set of parameters 
can be crucial for speeding up the convergence and achieving good accuracy in the ground state description (see, for instance, Fig. \ref{fig:pairing-nn}). 
\item The iterative procedure is stopped at a certain step when the maximal gradient Eq~(\ref{eq:gradient}) or the energy difference between the last few iterations is lower than a threshold ${\cal E}_{\rm ADAPT}$. 
\end{enumerate}

One advantage of the ADAPT-VQE technique is its flexibility in deciding operators of the pool. This flexibility was originally used to reduce the circuit depth when building the variational ansatz \cite{Gri19}. When a large set of operators is considered, the adaptive design of ansatz becomes a complicated combinatorial problem, and the method's success is often a balance between the pool size and the capability of optimization over a large set of parameters. Not surprisingly, after the original work of Ref. \cite{Gri19}, different strategies for refining the ADAPT-VQE technique have been explored. This includes the use of operators that can be built on quantum computers at low cost \cite{Yor22}, the reduction of the pool size by performing linear combination of pool operators based on physical arguments \cite{Sma21}, the possibility to perform several steps in one step as proposed in the TETRIS technique \cite{Ana24}, the adoption or rejection of operators breaking the physical problems' symmetries \cite{Rom22,Ber23}, and the optimization through information theory arguments \cite{Zha21}. Several open source codes are available \cite{Code1,Code2,Code3,Code4}. Some recently proposed techniques have also been tested in the present work and will be discussed below. 

\subsection{Illustration of the ADAPT-VQE technique to the 
particle-like pairing problem}
\label{sec:particlelike}

As a first illustration of the power of the ADAPT-VQE approach, we consider the case of a small superfluid system with only one type of Cooper pairs. This case corresponds to the simplest situation (case 1) reported in table \ref{tab:Hvariants}. Only pairs between protons or between neutrons can be formed, leading to two disconnected superfluids. For the sake of simplicity, we
consider here one of the two superfluids (namely, keeping only one particle species) and rewrite the Hamiltonian simply as:
\begin{eqnarray}
H &=& \sum_{p=1}^{n_B} \varepsilon_p N_p - g \sum_{p, q=1}^{n_B} P^\dagger_p P_q ,
\label{eq:Hnn}
\end{eqnarray}
where $N_p = a^\dagger_p a_p + a^\dagger_{\bar p} a_{\bar{p}}$ is the pair occupation operator, and where we assumed simply $g_V(-1) =g_V(+1) = g$. Here, 
$(p,\bar p)$ stands for pairs of neutrons (or protons) only. It is worth mentioning that today's
applications of the nuclear energy density functional theories usually assume only particle-like pairing, and the Hamiltonian (\ref{eq:Hnn}) can be seen as a simplified case 
where only the self-consistent effect of the pairing on the mean-field is neglected.

The treatment of particle-like pairing problems has already been extensively investigated on quantum computers using the QPE algorithm \cite{Ovr03,Ovr07}. Most recent studies focused on preparing accurate variational ansatz 
employing the SB/SR strategy \cite{Lac20}.  Specifically, using quantum ansatzes inspired by the BCS theory where wave-functions break the $U(1)$ symmetry, important progress has been made to prepare and use 
particle number projected ansatzes. Several techniques have been 
proposed to perform the symmetry restoration: symmetry restoration by phase estimation \cite{Lac20,Rui22}, direct classical post-processing \cite{Kha21} using, for instance, the classical shadow approach \cite{Hua20, Rui24}, purification by quantum oracles \cite{Rui23}, or direct construction of the projected state in the quantum circuit \cite{Kha23}.

\begin{figure}[htbp!]
    \centering
   \includegraphics[width = 0.8 \linewidth] {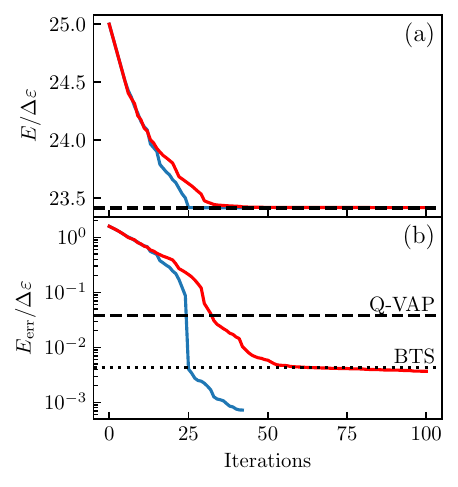}
    \caption{Panel a: Convergence of the ADAPT-VQE approach for the particle-like pairing problem as a function of the iteration number where the operator pool is defined with equation (\ref{eq:ab}). Here, the case of $N=10$ particles on $n_B=10$ doubly degenerated equidistant levels with single-particle energies 
    $\varepsilon_i = i \Delta \varepsilon $ with $i=0,1\ldots,N-1$ and $g/\Delta \varepsilon = 1$ is considered.
    The blue curve corresponds to the full procedure described in section \ref{sec:adaptgen}, while the red curve is the result obtained when omitting the readjustment of all parameters $(\theta_1, \cdots , \theta_n)$ at step $n$, i.e. only $\theta_n$ is adjusted at step $n$. The horizontal dashed line is the exact GS energy obtained by direct Hamiltonian diagonalization on a classical computer.  
    Panel b:  Error relative to the ground state energy in log scale. For comparison, we also display the results of the Q-VAP approach \cite{Rui22} based on a symmetry breaking/symmetry restoration strategy. The Q-VAP result (black dashed line)
    are those reported in Ref. \cite{Kha21}, note that in this reference, the Q-VAP approach is labelled by AGP 
    (Antisymmetric Geminal Power). The horizontal dotted line labelled by the BTS state corresponds to the result obtained in Ref. \cite{Kha23} by generalizing the AGP state using binary decision trees. }
    \label{fig:pairing-nn}
\end{figure}

We illustrate here that the ADAPT-VQE method can be competitive in solving the standard pairing problem without necessarily requiring the breaking of the $U(1)$ symmetry. Since most recent applications are restricted to even systems with seniority
zero, which allows to encode each pair on one qubit only, we use this simplified encoding here for Eq~(\ref{eq:Hnn}). Note that this simplification will not be used when considering the full Hamiltonian (\ref{eq:hamilNP}). The pair-to-qubit mapping is extensively discussed in the literature \cite{Kha21,Rui22}. The corresponding Hamiltonian written in terms of Pauli matrices reads:
\begin{eqnarray}
H = \sum_{p=0}^{N-1} (\varepsilon_{p}-g/2) \left[ 1 - Z_p \right] 
- \frac{g}{2} \sum_{p>q} \left[ X_p X_q + Y_p Y_q \right].
\label{eq:Hpairencoding}
\end{eqnarray} 
The matrices $(X_p,Y_p,Z_p)$ are the three standard Pauli matrices associated with the qubit $p$.
We see that, within the Hamiltonian restricted to seniority zero, 
using the simplified pair-to-qubit encoding gives a Hamiltonian that is a quadratic polynomial of the Pauli matrices. 
A pool of operators that is expected always to converge to the exact ground state can be guessed for this specific case. 
Indeed, a possible method to transform one state into another state 
in the qubit register basis is to apply sequentially a set of Givens rotations \cite{Nie00}. Since here we consider the restricted subspace of states having seniority zero, we can further restrict the Givens rotations such that the seniority is 
preserved. The operators that generate these specific Givens rotations for $p < q$ are:
\begin{eqnarray}
\displaystyle 
A_{pq} = i (P^\dagger_{p}P_q - P^\dagger_q P_p) = \frac{1}{2} (X_p Y_q - Y_p X_q).
\label{eq:ab}
\end{eqnarray} 
In total, assuming that we have $N$ 
doubly degenerated levels, the problem is encoded on $N$ qubits, and the number of operators in the pool is $N(N - 1)/2$. 

Note that this pool of operators can also be constructed from the two-body terms appearing in the Hamiltonian (\ref{eq:Hnn}).
Such a pool, inspired by the different sets of operators appearing in the Hamiltonian can be seen as a natural choice out of the adiabatic Gell-Mann and Low theorem \cite{Fet71}. This theorem states that starting from 
the ground state of an initial Hamiltonian $H_0$ (say, the one-body part in Eq~(\ref{eq:hamilNP})), we can evolve the state using a real-time propagator in which $H_0$ is adiabatically transformed into the final Hamiltonian $H_f$ and in the end, the final state should correspond to the ground state of $H_f$. Using Trotter-Suzuki decomposition to perform the adiabatic evolution shows that 
this evolution can be written as a product of short-time $\Delta \tau$ propagators $e^{-i \Delta \tau G_\alpha}$, where $G_\alpha$ are operators from $H$.
However, the brute-force application of the adiabatic theorem might require infinitely small time steps to converge. 
Replacing the $\Delta \tau$ by a set of free parameters allows a faster and possibly non-adiabatic evolution to the ground state. One example is the Variational Hamiltonian Ansatz (VHA) \cite{Wec15,Rei19,Ans21}.

Here, we do not use the VHA method itself but take inspiration from it to build a pool, such as Eq~(\ref{eq:ab}), which we call hereafter the H-pool. 
An illustration of the H-pool's performance on the simplified case of particle-like pairing is shown in Fig. \ref{fig:pairing-nn}. 
The optimizer used in this work is the Broyden-Fletcher-Goldfard-Shanno (BFGS) optimizer provided by \verb|SciPy|. The stopping criterion is reached if the difference in optimal energy of the last two iterations is below the threshold ${\cal E}_{\rm ADAPT}/\Delta \varepsilon = 10^{-6}$.
Two ADAPT-VQE calculations are shown, one where the full protocol discussed in section \ref{sec:adaptgen} is followed (blue solid line) 
and a second simpler calculation where we only adjust one parameter after the other, i.e. without re-optimizing the full set of parameters 
at step 3 (red solid line). For comparison, we show the result of the quantum variation after projection (Q-VAP) method that was recently 
adapted to quantum computers in Refs. \cite{Lac20,Kha21,Rui22}. The Q-VAP method can be regarded today as the state-of-the-art approach on classical computers and is based on a symmetry breaking/symmetry restoration technique. We also compare the results with the recent generalization of the Q-VAP ansatz proposed in Ref. \cite{Kha23} that is based on the Binary-Tree-State (BTS). 
We see that both calculations based on the ADAPT-VQE approach can converge to the ground state energy with rather high 
accuracy, comparable to that of the best existing ansatz (BTS case) for the pairing problem. In short, with the full re-optimization, ADAPT-VQE converges with much less parameters than in the case where only one parameter is adjusted per iteration, and outperforms the other approaches in searching the exact ground state. 

\section{Application of the ADAPT-VQE approach to the neutron-proton pairing problem}
\label{sec:adaptappliednp}

The illustrative example given above clearly points out that ADAPT-VQE  
can be a powerful approach to treating small superfluid systems. Note that this was also a conclusion 
of Ref. \cite{Bai24} where this approach was applied to the Agassi model that already includes pairing correlations.
We now consider the full Hamiltonian (\ref{eq:hamilNP}) where particle-like and particle-unlike pairing might coexist.  
The operator pool is a key ingredient of the approach. Specifically, the convergence of the approach is a balance between
(i) the ability of the operators to generate a sufficiently large subset of quantum ansatz that can approach the real ground state of the problem, a property that we call the expressive power of the operators pool, and (ii) the number of iterations required to achieve the convergence. On one hand, adding more operators to the pool should normally enhance the expressive power\footnote{Provided these new operators are linearly independent from the elements of the dynamical Lie algebra associated to the current pool.}. But on the other hand, it might complicate the optimization/combinatorics problem during the descent to the ground state and might, ultimately, prevent proper convergence.

\subsection{Encoding the spin-isospin pairing problem on qubits}

Until now, only one rather specific limit of the Hamiltonian (\ref{eq:hamilNP}), the one with seniority zero states, has been considered (case in \ref{tab:Hvariants}). The pair-to-qubit 
encoding was applied since we restricted the example to an even number 
of particles without pair breaking. In a general spin-isospin problem, because of the coexistence of several types of pairs, the pair-to-qubit encoding is not straightforward anymore in all situations. Even in the particle-like pairing, a more
general encoding should be employed if we are interested in non-zero seniorities or odd systems. The fermion-to-qubit Jordan-Wigner mapping \cite{Jor28,Lie61} will be suitable in these cases, as done in Refs. \cite{Ovr07, Lac20}. Therefore, we use the general Jordan-Wigner transformation (JWT) to encode the problem. The first step is to organize particles into a linear chain of fermions. This is done by grouping them into spin-isospin blocks of single-particle states, labelled by $i=1, \cdots , n_B$, and each block contains $4$ degenerated single-particle states.  The following ordering of these states is used: $[n \uparrow, n \downarrow, p \uparrow, p \downarrow]_i$. 
For instance, the state where $(n\uparrow , p \downarrow)$ are occupied in block $i$ corresponds to $| 1001 \rangle_i$.
\begin{figure}[htbp]
    \centering
    \includegraphics[width=\linewidth]{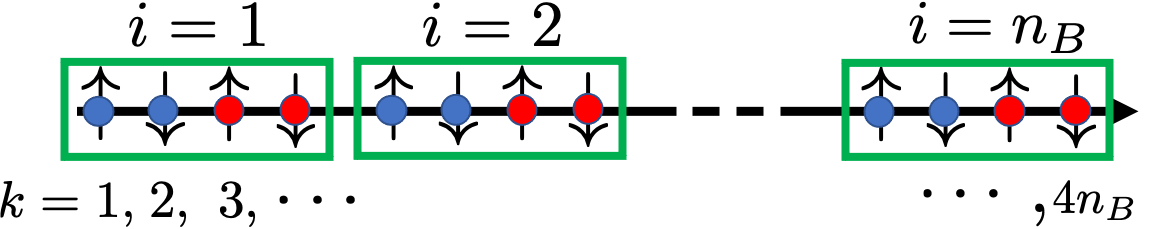}
    \caption{Schematic illustration of the single-particle ordering and mapping of the neutron-proton problem as a linear chain of fermions. The 
    neutrons (resp. protons) single-particle states are represented by blue (resp. red) circles. In each spin-isospin block $i$, the ordering is $[n \uparrow, n \downarrow, p \uparrow, p \downarrow]_i$. 
    The creation operators for protons and neutrons are then generically written as $a^\dagger_k$ with $k=1, \cdots, 4n_B$, where $n_B$ denotes the number of blocks (see text for further details on the mapping). 
    In the Jordan-Wigner fermion-to-qubit mapping, one qubit is associated with each single-particle state, leading to $q= 4n_B$ qubits. In this work, we use the little-endian convention. For instance, the state where 2 neutrons and 2 protons occupy the leftmost block corresponds to the state $| 0\cdots 0 1111 \rangle$. }
    \label{fig:enter-label}
\end{figure}

A schematic illustration of the ordering of the single-particle states is given in Fig. \ref{fig:enter-label}. From now on, we use the generic notation $a^\dagger_k$ for creation operators. Sometimes, it's more natural to locate a state with block index $i=1,\cdots, n_B$ and spin-isospin index $l=1,\cdots, 4$:
\begin{eqnarray}
    [a^\dagger_{i,1},a^\dagger_{i,2},a^\dagger_{i,3},a^\dagger_{i,4}] \equiv [\nu^\dagger_{i},\nu^\dagger_{\bar i},\pi^\dagger_{i},\pi^\dagger_{\bar i}] ,  \label{eq:map}
\end{eqnarray}
and then we assign to each $(i,l)$ a global index $k$ such that:
\begin{eqnarray}
    k = 4 (i-1) + l , \label{eq:ikl}
\end{eqnarray}
resulting in the short-hand notations $a^\dagger_k \equiv a^\dagger_{i,l}$.
Then, using the JWT, the qubit $k$ is linked to the state $k$. And we have the JWT correspondence: 
 \begin{eqnarray}
a^\dagger_k \longrightarrow  Q^+_k \otimes Z^{<}_{k-1}, ~~
a_k \longrightarrow  Q_k \otimes Z^{<}_{k-1},
\end{eqnarray}
with the convention:
\begin{eqnarray}
Q^{\pm}_k = \frac{1}{2} \left( X_k \mp i Y_k \right),  \label{eq:qpm}
\end{eqnarray}   
and $ Z^{<}_{k-1} =  \bigotimes_{j=1}^{k-1} (-Z_j)$. Here $(X_k, Y_k, Z_k)$ are the Pauli matrices of each qubit. 

With this encoding, the one-body terms of the Hamiltonian are easily obtained:
\begin{eqnarray}
    a^\dagger_k a_k &=& Q^+_k Q^-_k = \frac{1}{2}\left(I_k - Z_k\right), \nonumber
\end{eqnarray}
where $I_k$ is the identity matrix for the qubit $k$. The form of the two-body contribution to the Hamiltonian can be obtained from the expressions of the different pair operators $P^\dagger_{S_z,i}$ and 
$D^\dagger_{T_z,i}$ entering in Eq. (\ref{eq:hamilNP}-\ref{eq:pairoperators}). These expressions are given explicitly for the $6$ different $(S,S_z; T, T_z)$ channels in Table \ref{tab:pair}. In this table, we give the expressions of all pair creation operators in the 
$i^\text{th}$ block in terms of Pauli matrices and denote the obtained Pauli strings by ${\cal P}^\dagger_i(\alpha)$ with $\alpha=1, \cdots , 6$.   

The Hamiltonian acting on the qubit register, denoted by $H^{\rm JWT}$, is 
given by: 
\begin{eqnarray}
    H^{\rm JWT} = \frac{1}{2} \sum_{k=1}^{4n_B} {\varepsilon_k} \left[ I_k - Z_k \right] 
    -\sum_{\alpha=1}^{6} g_\alpha \sum_{i,j=1}^{n_B}  {\cal P}^\dagger_i(\alpha){\cal P}_j(\alpha). \label{eq:Hjwt} 
\end{eqnarray}
 The constants $g_\alpha$ are just $g_V(T_z)$ and $g_S(S_z)$ relabelled under the convention in Tab~{\ref{tab:pair}}. 
\begin{table}
    \centering
    \begin{tabular}{|c|c|c|l|}
    \hline \hline
     $(S,S_z)$ & $(T,T_z)$ & $\alpha$ & Qubit operators ${\cal P}^\dagger_{i}(\alpha)$ using JWT\\
     \hline
       "       &  (1,1)    & 1        & $Q^+_{i,1} Q^+_{i,2}$                                        \\
    (0,0)      &  (1,0)    & 2        & $(Q^+_{i,2}  Q^+_{i,3}-Q^+_{i,1} Z_{i,2} Z_{i,3} Q^+_{i,4})/\sqrt{2}$ \\
       "       &  (1,-1)   & 3        & $Q^+_{i,3}Q^+_{i,4}$                                        \\
    \hline 
    (1,1)      &  "        & 4        & $Q^+_{i,1} Z_{i,2} Q^+_{i,3}$ \\
    (1,0)      &  (0,0)    & 5        & $(Q^+_{i,2} Q^+_{i,3}+Q^+_{i,1} Z_{i,2} Z_{i,3} Q^+_{i,4})/\sqrt{2}$ \\
    (1,-1)     &  "        & 6        & $Q^+_{i,2} Z_{i,3} Q^+_{i,4}$\\
     \hline 
    \end{tabular}
    \caption{Expressions of the six pair operators $\{ P^\dagger_{S_z, i}\}_{S_z = -1,0,1}$ and $\{ D^\dagger_{T_z, i}\}_{T_z = -1,0,1}$ entering in the Hamiltonian (\ref{eq:hamilNP}) as a sum of Pauli strings using the JWT method and the ordering of states shown in Fig. \ref{fig:enter-label}. With a relabelling of subscript (\ref{eq:ikl}), $Q^+_{i,l}= Q^+_k$ where $k$ denotes the global index of a qubit/state on the linear chain.  
Useful relations that were used to simplify the expressions are 
$Z^2_m= I_m$, $Q^+_m Z_m = Q^+_m$ and $Z_mQ^+_m = -Q^+_m$. Note that we follow the little-endian convention in qubits ordering.}
    \label{tab:pair}
\end{table}

\subsection{Different pools of operators used for the proton-neutron pairing problem}
\label{sec:pools}

Three different pools of operators are used, varying in size and/or complexity to implement on a quantum computer:     
\begin{itemize}
    \item {\bf The Hamiltonian operators pool:} (H-pool) The first set of operators is a direct generalization of the pool used in section \ref{sec:particlelike}, and consists simply in using all operators that can be directly identified in the Hamiltonian after JWT, and that can be used as a generator of transformations. For the neutron-proton Hamiltonian, the pool consists of the set of operators 
    $\{Z_k\}_{k=1,\cdots,4n_B}$, complemented by the set of operators $\{ {\cal A}_{ij}(\alpha)
    \}_{i\le j=1,\cdots, n_B;\alpha=1,\cdots,6}$. The operators ${\cal A}_{ij}(\alpha)$
    generalize the ones introduced in Eqs. (\ref{eq:ab})
    to account for the larger number of pairing channels when considering all possible spin-isospin pairs (for $i < j$, and $\alpha=1,\cdots, 6$):
    \begin{eqnarray}
    \displaystyle {\cal A}_{ij}(\alpha) &=& i \left[{\cal P}^\dagger_{i}(\alpha) {\cal P}_{j}(\alpha) - {\cal P}^\dagger_{j}(\alpha) {\cal P}_{i}(\alpha)\right]. 
    \label{eq:abtotal}
    \end{eqnarray} 
    For a system of $n_B$ spin-isospin blocks, i.e. described on $4n_B$ qubits, the total number of operators in the pool scales as 
    $3n_B(n_B-1)$.
    The H-pool can be interpreted as a restricted class of Unitary Coupled Clusters technique based on single and double (UCCSD) fermionic excitation operators. 
    The restriction stems from the fact that the double excitation operators used in the evolution $\{ {\cal P}^\dagger_{i}(\alpha) {\cal P}_{j}(\alpha) \}$ will only change 
    the mixing of states within a specific channel, and each operator is block diagonal where each block corresponds to a given seniority.

    \item {\bf The qubit excitation pool \cite{Yor21}:} (QEB-Pool) One of the pools that are currently extensively used, especially in quantum chemistry, is UCCSD-inspired single and double qubits excitation operators. This pool is obtained from the equivalent fermionic excitation operators with fermion-to-qubit encoding but omits all the Pauli-$Z$ operators resulting from JWT \cite{Yor21}. The class of single and double qubits excitation operators is then given by (for $i<j<k<l$): 
    \begin{equation}
    \left\{
      \begin{array}{lll}
      T^{(1)}_{i j}&\equiv& (Q^+_{i}Q_{j}-Q^+_{j}Q_{i}) = \displaystyle\frac{i}{2}(X_{i}Y_{j}-Y_{i}X_{j}),\\ 
      \\
      T^{(2)}_{i j k l}&\equiv& Q^+_{i}Q^+_{j} Q_k Q_l - Q^+_{l}Q^+_{k} Q_j Q_i \\
      &=&  \displaystyle\frac{i}{8}(X_{i}Y_{j}X_{k}X_{l}+Y_{i}X_{j}X_{k}X_{l} \\ 
     &\displaystyle &+ Y_{i}Y_{j}Y_{k}X_{l}+Y_{i}Y_{j}X_{k}Y_{l}-X_{i}X_{j}Y_{k}X_{l} \\
     &\displaystyle &-X_{i}X_{j}X_{k}Y_{l}-Y_{i}X_{j}Y_{k}Y_{l}-X_{i}Y_{j}Y_{k}Y_{l}). 
  \end{array}
  \right.
  \label{eq:qeb}
    \end{equation}
    The associated circuits for these operator pools are given in Ref. \cite{Yor21}. Compared to the original UCCSD approach, it is advantageous in reducing the operator non-locality on the qubit register while preserving the total Hamming weight/particle number of the state, namely the number of $1$'s in the state. 
    Compared to the H-pool, some operators in Eq~(\ref{eq:qeb}) can mix different channels and states with different seniorities.
    Noteworthily, while the total particle number is conserved, the neutron and proton number might change simultaneously.
    
    \item {\bf The Qubit-pool \cite{Tan21}:} (Qubit-pool) A further simplification possible for the QEB-pool is to consider
    directly the qubit operators appearing in $T^{(2)}_{i j k l}$ in Eq~(\ref{eq:qeb}) , i.e.,  
    we also tested the so-called Qubit-pool, defined as the set of operators:
     \begin{eqnarray}
         \{i X_i Y_j , \; i X_{i}X_{j}X_{k}Y_{l},\; i Y_{i}Y_{j}Y_{k}X_{l}\}.
         \label{eq:qpool}
     \end{eqnarray}
     One motivation of the simplification is that the unitary operations built from these generators can be implemented by staircase circuits \cite{Nie00}, and are thus usually more hardware-efficient than linear combinations of these qubit operators. In terms of symmetry, only the qubit parity is preserved by Eq~(\ref{eq:qpool}). However, states with different Hamming weights can be mixed within a given parity block. When treating many-body problems, this implies that the $U(1)$ symmetry associated 
     with the total particle number conservation might be broken. This aspect will be further discussed in the illustration in Sec~{\ref{sec:systematic}}. 
\end{itemize}

A summary of the symmetries conserved and broken by the different pools is given in Table \ref{tab:pools_symmetries}. It is worth remarking that, by using the dynamical Lie algebra arguments developed in \cite{Ana22}, it can be shown that all three pools implement evolutions that can in principle reach the ground state, provided the initial state is suitably chosen. As our studies illustrate, this does not guarantee that the numerical solution of the variational problem indeed succeeds in reaching the GS, for a variety of reasons we expose below. 

\begin{table}[ht]
    \centering
    \begin{tabular}{|c|c|c|c|}
    \hline \hline
         & Particle number & Seniority & Parity\\
         \hline
      H-pool     & \checkmark & \checkmark & \checkmark\\
      QEB-pool   & \checkmark & $\times$   & \checkmark\\
      Qubit-pool & $\times$   & $\times$   & \checkmark\\ 
      \hline
      \end{tabular}
    \caption{Schematic view of the different symmetries that are preserved (with check mark) or might be broken (cross) during the ADAPT-VQE iterations for the different pools considered here. 
    Note that here ``Particle number'' and ``Parity'' refers to the total system particle number, i.e. the sum of neutron and proton numbers, and total parity. For instance, the QEB-pool, while preserving the total particle number, can lead to varying proton and neutron numbers during the ADAPT-VQE iterations. Only the H-pool preserves the particles number and parity for each particle species.}
    \label{tab:pools_symmetries}
\end{table}

\begin{figure}[htbp]
    \centering
\includegraphics[width=\linewidth]{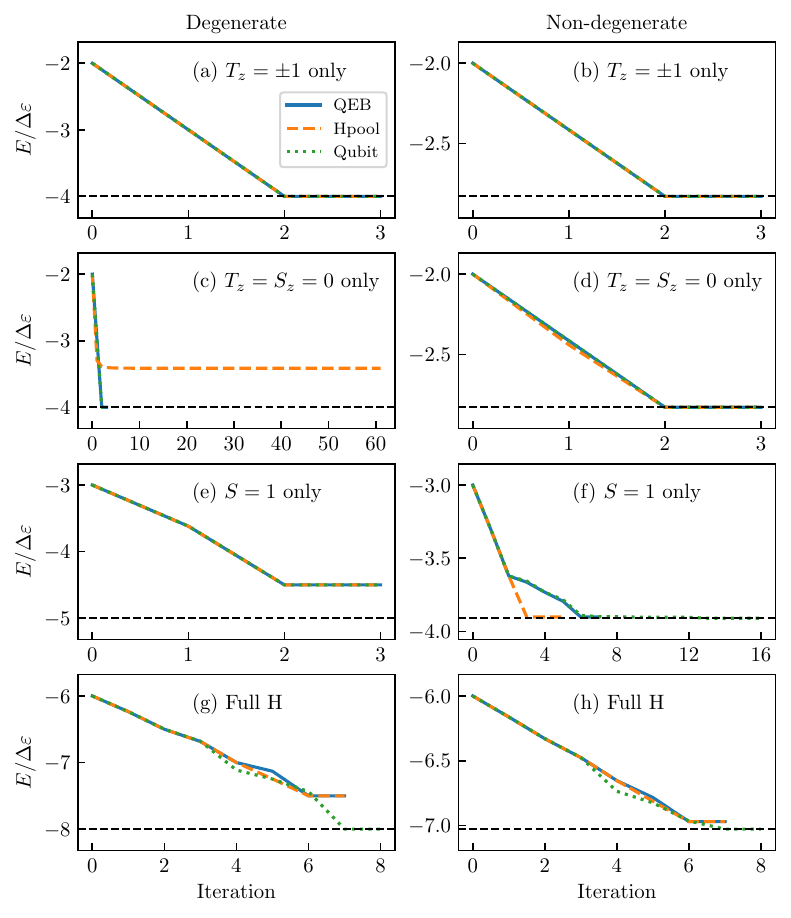}
    \caption{
    Evolution of the energies of the intermediate states optimized during the ADAPT-VQE iterations using the three different pools: H-pool (orange dashed line), QEB-pool (blue solid line), 
    and Qubit-pool (green dotted line), for the case of $2$ protons and $2$ neutrons on 8 qubits (2 spin-isospin blocks). The degenerate (left) and non-degenerate (right) spin-isospin blocks 
    are systematically shown for the 4 Hamiltonian cases given in table $\ref{tab:Hvariants}$, from top to bottom. 
    For the degenerate case, all energies are set to $0$, while, for the non-degenerate case, 
    the energies within two blocks are respectively $0$ and $\Delta \varepsilon=1$. All calculations 
    are performed with the same coupling constants $g= \Delta \varepsilon$ in the pairing channels that are not set to zero, and energies are given in $\Delta \varepsilon$ units. In all cases, the initial condition corresponds to a Slater determinant where 
    the 4 leftmost qubits in Fig. \ref{fig:enter-label} are set to $1$, i.e. $| 0000 1111  \rangle$. The black dashed line in each panel corresponds to the ground state energy.  
}
    \label{fig:adapt-4-on-2}
\end{figure}

\begin{figure}[htbp]
    \centering
\includegraphics[width=\linewidth]{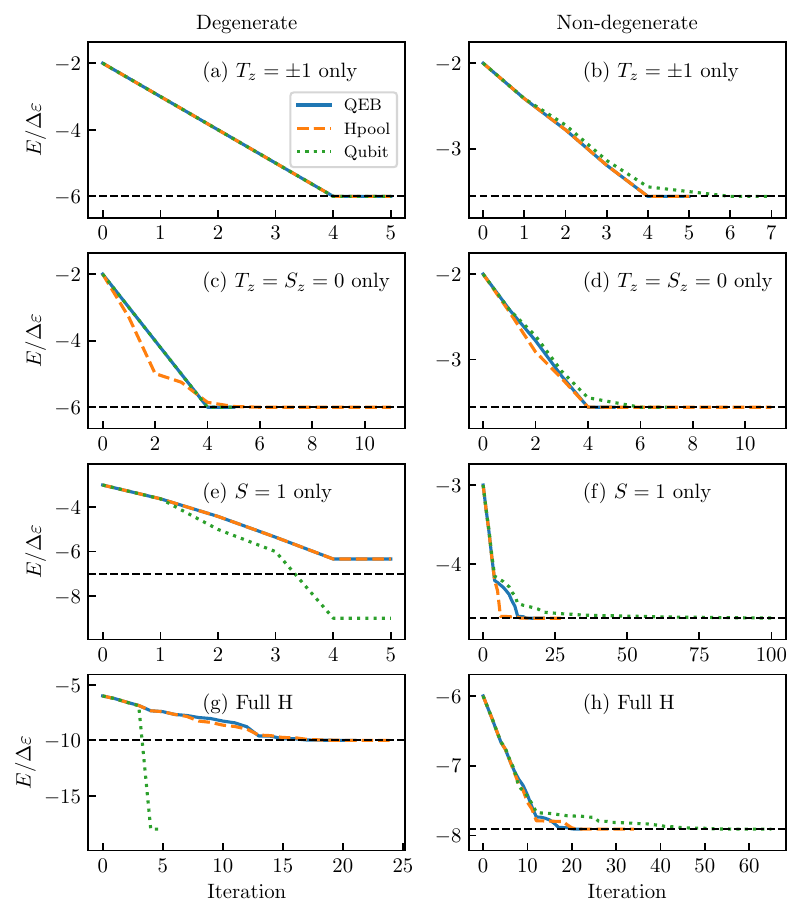}
    \caption{ Same as Fig. \ref{fig:adapt-4-on-2}, for the case of 2 protons and 2 neutrons on 12 qubits (3 spin-isospin blocks). For the non-degenerate case, the energies of the different spin-isospin blocks shown in Fig. \ref{fig:enter-label} are $0$, $\Delta \varepsilon$, and $2\Delta \varepsilon$ from left to right blocks. 
    In all cases, the initial condition is the same as the one used in Fig. \ref{fig:adapt-4-on-2}.  
    }
    \label{fig:adapt-4-on-3}
\end{figure}

\begin{figure}[htbp]
    \centering
\includegraphics[width=\linewidth]{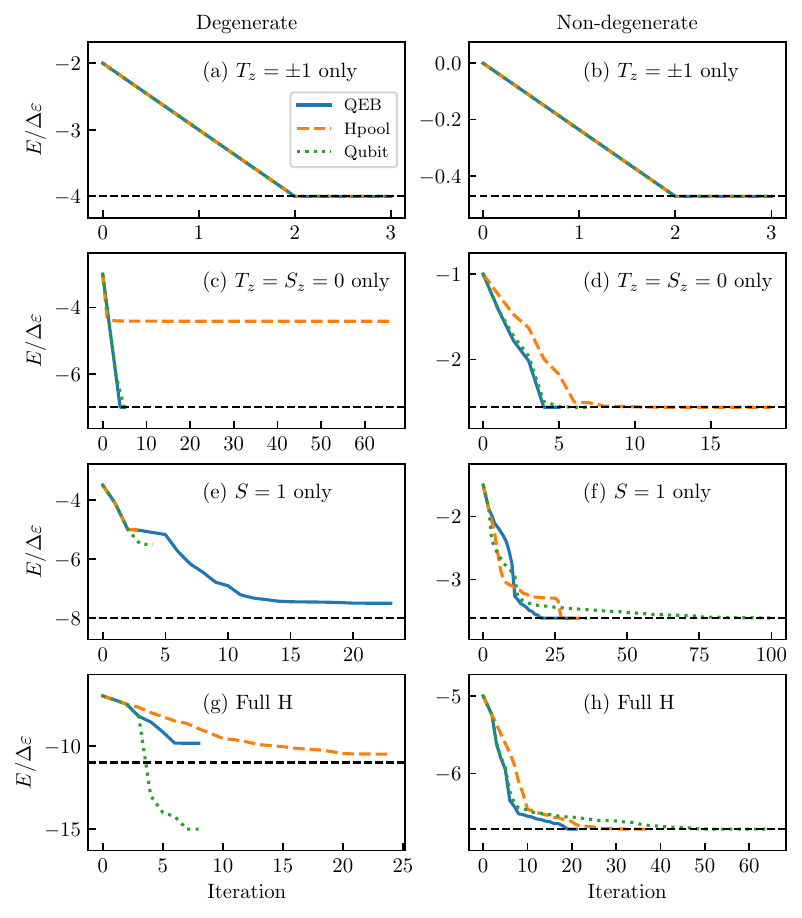}
    \caption{Same as Fig~{\ref{fig:adapt-4-on-3}}, but for 3 protons and 3 neutrons on 3 spin-isospin blocks starting from the initial state $\ket{000010011111}$.}
    \label{fig:adapt-6-on-3}
\end{figure}

\subsection{Systematic study of the pools' convergence towards ground states}
\label{sec:systematic}

Our objective here is to systematically investigate the expressive power of each pool, i.e., their capacity to express the ground state of Hamiltonians of the form (\ref{eq:hamilNP}). Specifically, we will benchmark the different ADAPT-VQE ansatz with various 
Hamiltonian listed in table \ref{tab:Hvariants}. In addition, we considered two types of single-particle levels: (1) the {\it degenerate} case where all blocks of 4 spin-isospin states are degenerated with single-particle energies set to zero, $\varepsilon_{i,n}=\varepsilon_{i,p}=0$, 
and (2) the {\it non-degenerate} case where the energy of 4 single particle states varies from one block to another ($\varepsilon_{i,n}=\varepsilon_{i,p} =\varepsilon _i$). Illustrations of results involving the aforementioned pools and different number of qubits are shown: 2 protons and 2 neutrons in $n_B = 2$ (Fig. \ref{fig:adapt-4-on-2}) and $n_B=3$ (Fig. \ref{fig:adapt-4-on-3}) blocks, as well as 3 protons and 3 neutrons in $n_B=3$ (Fig. \ref{fig:adapt-6-on-3}) blocks. 
The number of operators composing each pool is given in table \ref{tab:countop}. We clearly see from this table that the more symmetries are imposed, the fewer operators are used. And
reducing the number of operators in a pool is actually reducing the number of gradients to evaluate during the optimization. However, imposing more  symmetries on 
an operator is also accompanied by more operations in its circuit. We have, for instance, noted in practice that, with the same simulator, the execution time of the quantum circuits or the entire optimization is usually comparable between the H- and QEB-pools.  
\begin{table}
    \centering
    \begin{tabular}{|c|c|c|c|}
    \hline \hline
         & $n_B=2$ & $n_B=3$& $n_B=4$\\
         \hline
      H-pool     & 14 & 30   &52\\
      QEB-pool   &98  &561   &1940\\
      Qubit-pool & 140&  990 &3640\\ 
      \hline
      \end{tabular}
    \caption{Number of operators used for the cases with $n_B=2$, 3 and 4 spin-isospin blocks to obtain 
    the figures \ref{fig:adapt-4-on-2}, \ref{fig:adapt-4-on-3}, and $\ref{fig:adapt-6-on-3}$.}
    \label{tab:countop}
\end{table}

Several interesting features are observed in the Figs. \ref{fig:adapt-4-on-2}-\ref{fig:adapt-6-on-3}:
\begin{itemize}
    \item For all non-degenerate cases, we systematically observe a good convergence of all pools towards the ground state energies. Only in Fig. \ref{fig:adapt-4-on-2}-h, the ADAPT-VQE method gives a slightly higher energy. We also note in particular that the Qubit-pool always predicts the ground state energy to a good precision, but sometimes requires a larger number of iterations compared to other pools, as shown in the right panels of Fig. \ref{fig:adapt-4-on-3}. 
    \item For the degenerate counterparts, in some cases no specific convergence problem is observed. But we encounter pitfalls with certain combination of pools and Hamiltonian variants. The iteration might stop after reaching some plateau during the descent. Or, we see that the energy with Qubit-pool can drop drastically below the exact GS. 
\end{itemize}

We investigated the possible origin of the ADAPT-VQE non-convergence. In general, when the algorithm does not converge, tracing back 
the origin of the problem is rather complex due to the algorithm and the optimizer combining as a black box to select a sequence in the pools. One regular 
difficulty is the possible existence of the so-called barren plateau problem, that has been widely investigated 
and is largely discussed in the literature (see \cite{Lar22,Til22,Lar24} and Refs therein). We discuss below several difficulties that we identify and/or that might be the source of the problems we observe:
\begin{enumerate}

    \begin{figure*}[htbp]
  \centering
  \includegraphics*[width=0.3\linewidth]{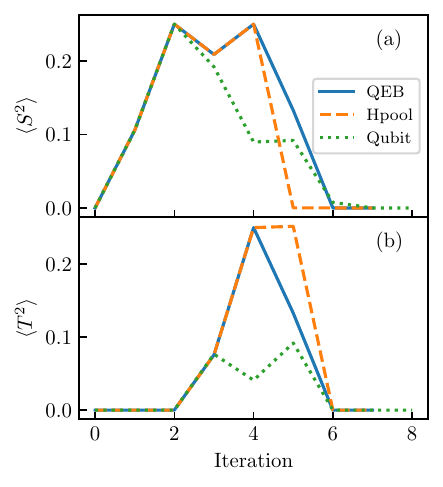}
 \includegraphics*[width=0.3\linewidth]{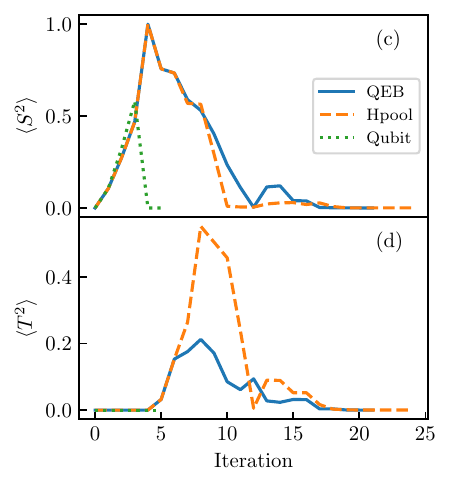}
 \includegraphics*[width=0.3\linewidth]{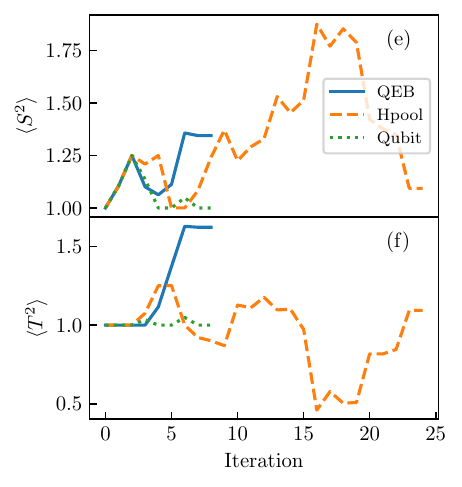}
  \caption{Evolutions of $\langle {\bf S}^2 \rangle$ and $\langle {\bf T}^2 \rangle$ during the ADAPT-VQE iterations for [Panels (a-b)] 2 protons and 2 neutrons on 2 spin-isospin blocks (Fig. \ref{fig:adapt-4-on-2}-g case), [Panels (c-d)] 2 protons and 2 neutrons on 3 blocks (Fig. \ref{fig:adapt-4-on-3}-g case), and [Panels (e-f)] 3 protons and 3 neutrons on 3 blocks (Fig. \ref{fig:adapt-6-on-3}-g case). We see that the expectation values of the total spin and total isospin might explore various values during the iterations. Still, when the energy converges to the ground-state energy, the broken symmetry might eventually be automatically restored, as shown in panels (a-d). Even if the particle number is not preserved in the Qubit-pool case, the final total spin and isospin might converge towards the expected values $S=T=0$ as expected in panels (c-d). In panels (e-f), whatever is the pool, 
  none of the final states converge to a physical state. Note that in this case both asymptotic $S(S+1)$ or $T(T+1)$ are equal to $1$. This is due to the fact that the final state does not end up 
  in a given $S$ or $T$ total spin block, but has non-zero amplitudes in different $S$ and $T$ symmetry blocks while having the same degenerated energies. 
  }
  \label{fig:T2S2}
\end{figure*}

    \item {\bf Spontaneous symmetry breaking:} In panels (e) and (g) of Fig. \ref{fig:adapt-4-on-3} or panel (g) of Fig. \ref{fig:adapt-6-on-3}, the energy obtained with the Qubit-pool suddenly becomes lower than the GS energy. This is due to the breaking of the 
    $U(1)$ symmetry associated with particle number conservation during the iteration. Said differently, during the optimization, there is a sudden change in the particle number after certain operator is applied to the state. The spontaneous symmetry breaking during the optimization might not always lead to such a dramatic effect in the energy. Indeed, in practice, we have observed that some symmetries of the Hamiltonian, like the total spin and/or total isospin symmetry, can be slightly broken. However, the ADAPT-VQE might still find the right solution as the broken symmetry is automatically restored during the iterative process (see Fig. \ref{fig:T2S2}). 
    A more catastrophic scenario happens in Fig. \ref{fig:adapt-4-on-3}-e and -g or Fig. \ref{fig:adapt-6-on-3}-g. The particle number symmetry is slightly broken at some intermediate steps, but this breaking is amplified at the next steps. Since all parameters are re-adjusted at each step, a significant change of energy can occur even in one step. This effect strongly jeopardizes results obtained from a pool that does not preserve the number of particles/Hamming weight of the state, as is the case for the Qubit pool here. Note that a pool preserving the $U(1)$ symmetries can still break other non-trivial symmetries. For example, in the degenerate case, the Hamiltonian commutes with both ${\bf T}^2$ and ${\bf S}^2$. 
    But for both the QEB- and H-Pool, even if the starting point is a proper eigenstate of these operators, we observe that the expectation value $\langle {\bf T}^2 \rangle$ and $\langle {\bf S}^2\rangle $ might vary across iterations. However, sometimes the effect on the energy is less dramatic than that from particle-number symmetry breaking and as a result, the algorithm still finds the correct path to the GS, restoring these symmetries automatically (see illustrations in Fig. \ref{fig:T2S2}).      
    Inspired by this, we have tested several methods to alleviate the symmetry violation if the pool of operators might spontaneously break it (see the next section). In \cite{Ber23}, a specific study of the advantages and drawbacks of using operators that eventually break symmetries was performed. In the work, they considered the possibility of breaking parity or total spin symmetries and did not observe any sudden jump because of particle number conservation. However, they also concluded that the number of iterations/parameters might counterbalance the gain in breaking symmetries to build an accurate approximation of the GS.
    
    \item {\bf Ground state degeneracy:} The degenerate case of uniform single particle energies 
    is a rather extreme problem. Indeed, in this case \cite{Pan69}, the ground states of all $(S,T)$ channels become degenerate. We took this example especially for benchmarking the performance of the algorithm.  Since the ADAPT-VQE is guided solely by the gradient of the energy, Eq. (\ref{eq:grad}), when some of the operators might break the Hamiltonian symmetry, different $(S,T)$ blocks in Hilbert space might be mixed 
    and the final wave-function might decompose onto various blocks. Still, the correct ground-state energy is found if the final state decomposes solely on GS in each $(S,T)$ block. However, avoiding the mixing of different symmetry blocks might be critical to extract physical quantities. 
    \item {\bf Dependence on the initial state:}
    We tested that the non-convergence observed in Figs. \ref{fig:adapt-4-on-2}-\ref{fig:adapt-6-on-3} in some cases  might eventually be cured 
    by preparing the initial state with some preliminary rotation. The origin of this effect is not always clear, but one might point out that a key ingredient of the pools is the "starters" determined partially by the initial states. Starters are the subset of pool operators that can induce a departure 
    from the initial states. This number-, state-, pool-, and Hamiltonian-dependence significantly affect the overall convergence \cite{Shk23}.
    The dependence on the initial state will be exploited in the section \ref{sec:randominitial} below.
    
    \item We finally mention that the H- and QEB-pools usually give similar convergence patterns. 
\end{enumerate}

Our conclusion for this systematic study is that the ADAPT-VQE can provide good 
results for the different pools in most situations. 
Because of the particle number non-conservation, the Qubit-pool is hard to control and might 
escape outside the physical wave-function space (that is, the one with the correct particle number), leading to undesirable 
minimization of the variational principle. 
For the other two pools, we generally observed that they give very similar, sometimes almost identical, results. In the next section, we propose some methods to improve the convergence of the ADAPT-VQE algorithm, focusing especially on the QEB-pool.

\section{Methods to control and/or improve the convergence}
\label{sec:improve}

We see in Figs. \ref{fig:adapt-4-on-2}, \ref{fig:adapt-4-on-3}, and \ref{fig:adapt-6-on-3} that the 
ADAPT-VQE can converge in almost all non-degenerate cases, whatever the pool is. For the degenerate cases, the ADAPT-VQE technique might have difficulties achieving good precisions on the GS's energy in many situations.    
During this work, we have developed several techniques that might, in some cases, improve the convergence. Some of the techniques introduced here are 
guided by methods employed to restore symmetries on a classical computer, while others are more empirical.

\subsection{Symmetries control during the optimization}
\label{sec:controlsymmetries}

The ADAPT-VQE technique is based on a pre-selected pool of operators. 
We observe in Fig. \ref{fig:adapt-4-on-2}-g that the Qubit-pool is the only pool that properly converges to the GS. In this case, breaking symmetries seems to help achieving convergence. On the other hand, by going out from the physical Hilbert space, it might also lead to completely erroneous results, as in Fig. \ref{fig:adapt-4-on-3}-g,  or \ref{fig:adapt-6-on-3}-g. Noteworthily, here, we can access the exact solutions and check that the GS is properly found. This will not be the case in future applications, and a final energy lower than the true GS energy might be hard to trace back.  

We tested two directions to control, at least to some extent, the symmetries during the energy minimization: the explicit projection of the 
trial ansatz or the addition of a penalty function. These techniques allow to better navigate through the Hilbert space of ansatzes, and sometimes help to reach lower energies. In most cases we studied, the full convergence to the GS is not achieved. For the sake of completeness, we still describe these techniques in this section.  

\subsubsection{Projection during the descent}
\label{sec:projection}

A natural method to ensure that symmetries are preserved during the iteration is to replace the trial wave-function 
Eq. (\ref{eq:trialadapt}) by its projection on the targeted symmetry sector. We denote by ${P}_{\cal S}$
the projector associated with a certain symmetry's expectation value, generically labelled by ${\cal S}$ below. A trial state vector that will never break this symmetry 
can be written as:
\begin{eqnarray}
    | \Psi'_n \rangle &=& {P}_{\cal S} | \Psi_n \rangle, \label{eq:proj}
\end{eqnarray}
where $| \Psi_n \rangle$ is the original iteratively updated trial state, and where the symmetry of interest could be, for instance, the
proton and neutron numbers, the parity, and/or the total spin or isospin. 
The implementation of projection techniques has been largely explored recently in the quantum computing context, and various techniques can be used to perform the projection 
\cite{Lac20,Siw21,Rui22,Lac23}.  
\begin{figure}[htbp]
  \raggedright
    \includegraphics*[width=0.45\textwidth]{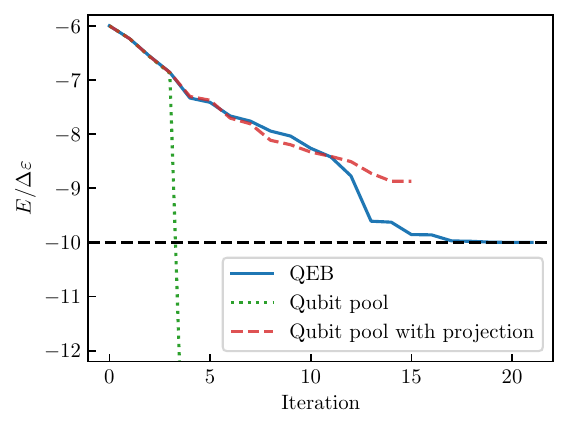}
  \caption{Same as Fig. \ref{fig:adapt-4-on-3}-g. The QEB- (blue solid line) and Qubit-  (green dotted line) pools are again displayed. 
  The new red dashed curve corresponds to the ADAPT-VQE results using the Qubit-pool with projection on $N=2$ neutrons and $Z=2$ protons as discussed 
  in section \ref{sec:projection} using Eq. (\ref{eq:proj}). }
  \label{fig:projection}
\end{figure}

We tried this technique to correct the behavior seen in Fig. \ref{fig:adapt-4-on-3}-g for the Qubit-pool. Specifically,
we used the state $P_{N=2} P_{Z=2} | \Psi_n \rangle$ 
to estimate the gradient and optimize the variational parameters, where $P_{N=2}$ (resp. $P_{Z=2}$) is the projector on
the neutron (resp. proton) number equal to $2$. Results of the method are illustrated in Fig. \ref{fig:projection}. 
As expected, the sudden jump in energy is suppressed and the  energy decrease becomes comparable to the QEB-pool case. However, the optimization turns out to stop 
before reaching the targeted energy, as it is sometimes observed for other pools that do not break the particle number symmetry.  
In addition, we mention that 
performing projections on a quantum computer is rather demanding regarding quantum resources, as discussed in Ref. \cite{Lac23}. Finally, projection might counterbalance the advantage of the ADAPT-VQE technique introduced 
specifically to reduce quantum computing resources. Note that the projection of particle number is one of the simplest projections. Projection on ${\bf S}^2$, 
or equivalently, on ${\bf T}^2$, as was made in Refs. \cite{Siw21,Rui24}, is more demanding since these operators are two-body operators. 

\begin{figure}[htbp]
    \centering
  \includegraphics*[width=\linewidth]{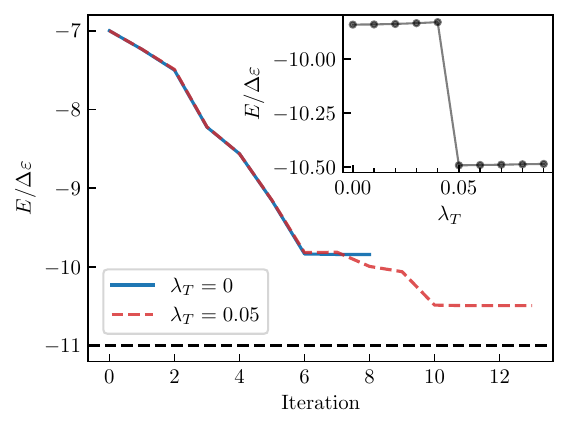}
  \caption{Illustration of the energies obtained during the iteration with the ADAPT-VQE using the original QEB approach without (blue solid line) or with 
  (red dot-dashed line) the penalty term introduced in Eq. (\ref{eq:penalty}) for the case of 3 protons and 3 neutrons on 3 blocks with an initial state 
  $\ket{000010011111}$. In the penalty case, the value $t=1$ is used, which corresponds to one of the degenerated GS. The inset shows the final energy obtained with different $\lambda_T$ values. We see that the final energy is rather stable over the range of explored $\lambda_T$. When larger values of $\lambda_T$ are explored (not shown), 
  the final energy suddenly jumps to a value close to or slightly higher compared to the case 
  without the addition of a penalty term, and the method breaks down.
  }
  \label{fig:penalty}
\end{figure}
\subsubsection{Penalty function method}
\label{sec:penalty}

Another possible approach to guide the ADAPT-VQE convergence towards a certain symmetry sector is to add one or several constraints during the energy descent. For instance, one can constrain the total isospin operator by considering the alternative Hamiltonian $H_c$ defined as 
\begin{eqnarray}
    H_c = H + \lambda_T \left[{\bf T}^2 - t (t+1) \right]^2,  \label{eq:penalty}
\end{eqnarray}
where $t$ is the targeted value of the symmetry sector under interest. We show in Fig. \ref{fig:penalty} an illustration of the gain 
in energy obtained by adding a penalty function for the case of 3 neutrons and 3 protons on 12 qubits (Fig. \ref{fig:adapt-6-on-3}-g case). This figure illustrates that adding a constraint 
might help reach lower energies when convergence is not fully achieved. We also see in this example that the asymptotic energy is still higher than the GS energy and that the ADAPT-VQE with constraint is insufficient to provide full convergence
in this case.   

\subsection{Other methods to improve convergence in ADAPT-VQE}

In the course of the present work, we have developed and tested several 
methods guided by the analyses of the different results shown in Fig. \ref{fig:adapt-4-on-2}, 
\ref{fig:adapt-4-on-3}, and \ref{fig:adapt-6-on-3}. Some of the methods are summarized below. Again, we concentrate on the degenerate case 
where difficulties in obtaining the GS energy are seen.

\subsubsection{Addition of operators/starters  to the pool}
\label{sec:starter}

Provided that the optimization is doable, one can add to the pool new operators that can provide new possibilities for jumping between the state at step $n$ and $n+1$. 
The interest of extending the ``link'' between different states was evidenced, for instance, in Ref. \cite{Shk23} for the very first step, where the notion of ``starters'' is introduced, i.e. the specific subsets of operators that allow to perform transitions from 
the initial state. The number of starters depends on the initial state. Given such a state, adding starters can help and/or accelerate the convergence.  
In practice, adding more and more operators might lead to a cumbersome optimization task, and we are left with minimizing the number of added operators to achieve convergence.
\begin{figure}[htbp]
  \raggedright
    \includegraphics*[width=0.43\textwidth]{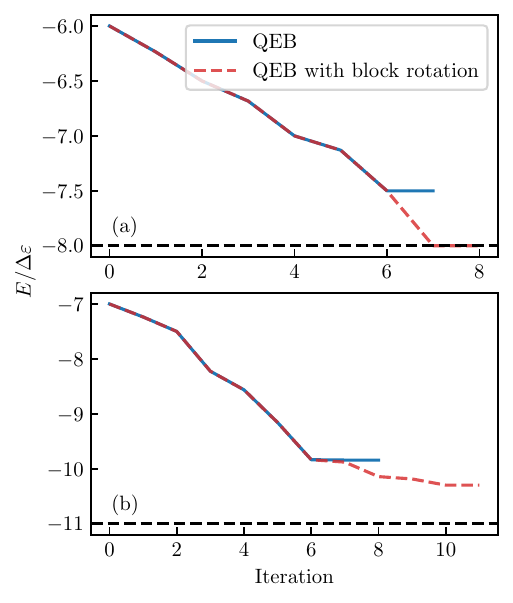}
  \caption{Results of ADAPT-VQE using the QEB-pool without (blue solid line) and with (red dashed line) the addition of the 
  block-rotation operator given by Eq. (\ref{eq:blockrotation}) to the pool. Both panels correspond to 2 protons and 2 neutrons 
  in a set of $2$ (a), or $3$ (b) degenerate blocks with the full Hamiltonian. The blue solid curves are the results presented in Fig. \ref{fig:adapt-4-on-2}-g and \ref{fig:adapt-6-on-3}-g in panels a and b, respectively.
 }
  \label{fig:qeb+blk-rot}
\end{figure}

As an illustration of the effect of adding operators, we consider again the case of 
Fig. \ref{fig:adapt-4-on-2}-g of 2 protons and 2 neutrons on 2 degenerate blocks with the full Hamiltonian. During the convergence, the energy for the QEB-pool gets stuck at an energy $0.5 \Delta \varepsilon$ above the exact GS energy. Analyzing the final state, we found that this stems 
from the fact that the pool cannot properly account for the block permutation invariance of the problem. The final state of the QEB-ADAPT-VQE approach has unbalanced amplitude for the 
state $|00001111 \rangle$ and $|1111 0000 \rangle$ in the final state. The former is the initial state used in Fig. \ref{fig:adapt-4-on-2}-g. Going from this state to the latter is not possible in one iteration, given that the QEB-pool is formed only by one-body and two-body transitions.  One can remedy this problem simply by adding to the pool the operator $T_4$, defined by:
\begin{eqnarray}
    T_4 &\equiv& i \Big( | 11110000 \rangle \langle 0000 1111 | \nonumber \\
    && - | 0000 1111 \rangle \langle 1111 0000 | \Big). \label{eq:blockrotation} 
\end{eqnarray}
This operator will be referred to as a block-rotation operator. For larger block numbers, one can generally define $n_B(n_B-1)/2$ such operators.  

In Fig. \ref{fig:qeb+blk-rot}, we give two illustrations of the effects of adding block rotation 
operators, respectively, for $2$ and $3$ blocks. For the $2$ blocks case (panel a), with a single operator 
added that also acts as a starter for the specific initial state, the new pool leads to 
proper convergence to the GS with a single additional iteration compared to the original QEB-pool. Unfortunately, adding such starters does not always lead to a perfect convergence to the GS. This is illustrated in Fig. \ref{fig:qeb+blk-rot}-b for 3 protons and 3 neutrons 
on 3 blocks. In this case, 3 block-rotation operators are added to the QEB-pool. With this addition, the full convergence to the GS is not achieved, but the few added operators help 
to lower the final energy.  

Noteworthily, the need or not to increase the number of operators in the pool is rather subtle. 
Indeed, with the very same initial state, the same Hamiltonian, and without the addition of any starter, the ADAPT-VQE approach can properly converge to the GS in the 3 blocks case 
(Fig. \ref{fig:adapt-4-on-3}-g), while it was not in the 2 blocks case (Fig. \ref{fig:adapt-4-on-2}-g). 

\begin{figure}[htbp]
  \raggedright
    \includegraphics*[width=0.43\textwidth]{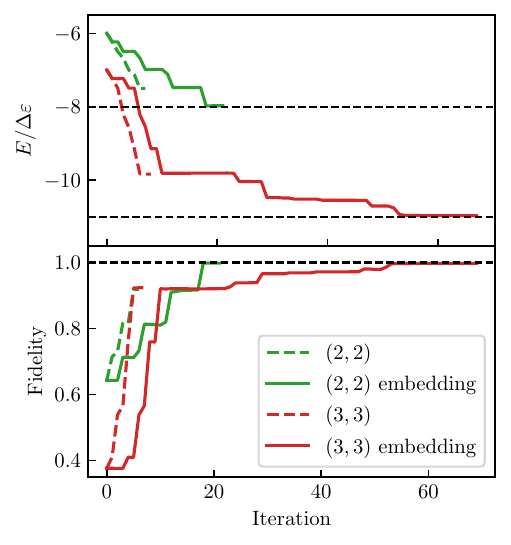}
\caption{
     Examples of energy (a) and fidelity (b) evolutions during the QEB-ADAPT-VQE iterations when the embedding technique is used to treat the fully degenerate case with the full Hamiltonian. The fidelity is defined as the square of the overlap between the ADAPT-VQE 
     state and the targeted ground state.  
     Two cases are illustrated: the case of 2 protons and 2 neutrons in a system 
     described on $n_B=2$ spin-isospin blocks (green) [Fig. {\ref{fig:adapt-4-on-2}}-g], and
     the case of 3 protons and 3 neutrons on $n_B=3$ blocks (red) [Fig~{\ref{fig:adapt-6-on-3}}-g]. An energy gap $E_g/ \Delta \varepsilon =20$ was used in both cases. Note that here the notation $(2,2)$ (resp. $(3,3)$) simply means that 2 (resp. $3$) particles of each type are considered on $2$ (resp. $3$) degenerated blocks. For the $n_B=3$ blocks, we embedded the problem using 4 blocks. For the sake of completeness, we give in table \ref{tab:countop} the associated number of operators in the pool.  
      }
  \label{fig:embedding}
\end{figure}

\subsubsection{Embedding degenerate problems into non-degenerate ones}

QEB-ADAPT-VQE performs generally well in non-degenerate cases, while, for several Hamiltonian 
cases listed in table \ref{tab:Hvariants}, we encountered difficulties to converge 
to the GS energy, regardless of the proton/neutron numbers. Based on the accuracy of the technique to describe the non-degenerate case, a   
possible strategy 
we explored is to transform a degenerate problem into a non-degenerate one. This is achieved in practice  by adding a set of 
ancillary qubits/blocks having single-particle energies that are different from the others. Explicitly, 
assuming a number $n_B$ of degenerate blocks where all single-particle energies are equals 
to $\varepsilon_1$ (possibly $\varepsilon_1 = 0$), we add an extra block of 4 qubits 
all associated to the energy $\varepsilon_1 + E_g$, i.e. 
\begin{eqnarray}
\varepsilon_k = \left\{
\begin{array}{ll}
\varepsilon_1, & \quad k=1,2,\ldots 4n_B \\
\varepsilon_1 + E_g, & \quad k=4n_B+1,\ldots, 4n_B+4
\end{array}
\right. ,
\label{eq:embedding}
\end{eqnarray}
where the new positive parameter $E_g$ is the energy gap between the system and the extra block. The total Hamiltonian is still given by Eq. (\ref{eq:hamilNP}) with parameters fixed for one of its variants given in table \ref{tab:Hvariants}. The last block acts as an environment on the system of interest formed by the first $n_B$ blocks, and one can decompose the total Hamiltonian as:
\begin{eqnarray}
    H_{n_B + 1} &=& H_{n_B} + H_{1_B} + H_{\rm Coup},
\end{eqnarray}
where $H_{n_B}$ stands for the Hamiltonian associated with the system assumed to be degenerated. $H_{1_B}$ is the Hamiltonian of the extra block, that reduces to the one-body terms of the 4 extra qubits. $H_{\rm Coup}$ is the part of the two-body term that 
couples the system with the extra block.  By performing such embedding, the system, labelled 
below by $S$, becomes an open quantum system coupled to a small environment labelled by $E$.

We now use the generic notation $| \Psi \rangle$ for a wave-function 
described on the full system + extra block. In this way, the system is embedded 
in a larger Hilbert space, and one can access its properties by tracing over the environment's degrees of freedom, i.e. 
\begin{eqnarray}
    \rho_{S} &=& {\rm Tr}_E \left( | \Psi \rangle\langle \Psi  |\right).
\end{eqnarray}
Noteworthily, since the system and environment become entangled due to the coupling, the reduced system density differs from a pure state density unless the coupling term 
$H_{\rm Coup}$ is set to zero. Here, we take $E_g$ as a free parameter that we can arbitrarily 
set to infinity. And as the energy gap increases, we expect that:
\begin{eqnarray}
    \lim_{E_g \rightarrow \infty} {\rm Tr}(H_{n_B}\rho_{S}) = E_{\rm GS}(n_B), \label{eq:partial}
\end{eqnarray}
with $E_{\rm GS}(n_B)$ the ground state energy associated to the system with $n_B$ blocks. 
Provided that the ADAPT-VQE approach converges for a moderate value of $E_g$, the method we propose is then to perform calculations with increasing values of $E_g$ in the non-degenerate case and 
deduce the degenerate case energy from the partial trace given by Eq. (\ref{eq:partial}). Notably, the partial trace can be avoided 
simply by computing $\langle \Psi | H_{n_B} | \Psi \rangle$ 
during the ADAPT-VQE iteration. We made extensive tests of the embedding technique and found that this approach is systematically able to cure the problem of convergence for the 
degenerate neutron-proton pairing problem, for all the variations of  Hamiltonian with a proper choice of $E_g$.  

Two illustrations of convergence patterns are shown in Fig. ~{\ref{fig:embedding}} for systems described on $2$ or $3$ blocks. The two cases are taken from the examples of section \ref{sec:systematic} where QEB could not converge to the correct 
energy. We see that the embedded approach not only leads to a perfect convergence towards the exact energy but it also gives a wave-function in high fidelity/overlap with the exact ground state wave-function. 
Among the methods we have tested to improve the ADAPT-VQE approach based on the QEB-pool, this approach seems to systematically lead to convergence. It should be noted that 
this technique also leads to a significant increase in the numerical cost since it not only 
increases the number of qubits but also increases the number of operators in the pool. Indeed, 
as shown in table \ref{tab:countop}, the pool size significantly increases 
from $n_B$ to $n_B+1$ blocks. Still, these operators 
are restricted to single and double excitation operators, not requiring any higher order excitation as in section \ref{sec:starter}.

\subsubsection{Randomization of initial states preparation}
\label{sec:randominitial}

As discussed in section \ref{sec:starter}, changing the initial state 
might change the number of starters and, ultimately, might improve and/or degrade
the convergence patterns. Choosing a specific initial state, as we did in previous 
examples, can lead to a biased conclusion about the quality of the selected operator pools. One way to avoid erroneous conclusions on this aspect is to consider 
a sufficiently large class of initial states and perform the ADAPT-VQE procedure
for each state. The set of initial states should be sufficiently simple 
to not induce too much additional gates for their preparation on a quantum circuit. 
\begin{figure}[htbp]
    \centering
\begin{tikzcd}[row sep = 0.2cm, column sep=0.2cm]
\ket{0} &	&	\qw	&	\gate[wires=2]{A(\theta_1,\phi_1)}	&	\qw	&	\gate[wires=2]{A(\theta_4,\phi_4)}	&	\qw	&	\qw	\\
\ket{0} &	&	\gate{X}	&	\ghost{A(\theta_1,\phi_1)}	&	\gate[wires=2]{A(\theta_3,\phi_1)}	&	\ghost{A(\theta_4,\phi_4)}	&	\gate[wires=2]{A(\theta_6,\phi_6)}	&	\qw	\\
\ket{0} &	&	\gate{X}	&	\gate[wires=2]{A(\theta_2,\phi_1)}	&	\ghost{A(\theta_3,\phi_1)}	&	\gate[wires=2]{A(\theta_5,\phi_5)}	&	\ghost{A(\theta_6,\phi_6)}	&	\qw	\\
\ket{0} &	&	\qw	&	\ghost{A(\theta_2,\phi_1)}	&	\qw	&	\ghost{A(\theta_2,\phi_2)}	&	\qw	&	\qw
\end{tikzcd}
    \caption{Illustration of a circuit performing a set of operations preserving 
    the neutron Hamming weight for the $n_B=2$ blocks case. In this figure, only 
    the qubits associated with neutrons ($k=1$, $2$, $5$ and $6$ in Fig. \ref{fig:enter-label}) are shown to simplify the circuit since the set of operations on the neutrons is independent of the one for protons. Similar operations are performed with different $(\theta_i, \phi_i)$ angles on the protons qubits. This figure presents a specific situation where the spin-down of the first block (resp. spin-up of the second block) is first occupied through the 
    $X$ operations. Then, specific operations through the $A$ gate are performed. This figure exactly corresponds to Fig~4 in \cite{Gar20}, and the $A$ 
    operators are building blocks performing mixing of the $| 01 \rangle$ and 
    $|10 \rangle$ states on the two qubits it applies. The $A$ operator 
    is defined by the matrix (2) of Ref. \cite{Gar20}, together with the associated circuit given in the same reference.  }
    \label{fig:gard}
\end{figure}
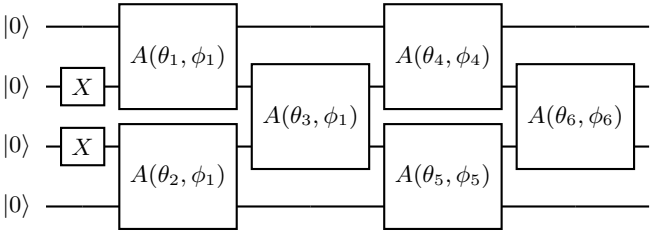

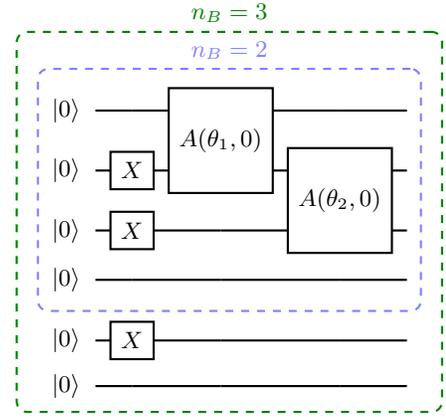
\begin{figure}[htbp]
    \centering
\begin{tikzcd}[row sep = 0.2cm, column sep=0.2cm]
    \gategroup[7,steps=6,style={dashed,rounded
    corners,color=green!50!black, inner
    xsep=10pt, inner ysep = 10pt, yshift=8pt},background,label style={color=green!50!black,label
    position=above,anchor=south,yshift=-0.2cm}]{{$n_B=3$}}
    \gategroup[4,steps=6,style={dashed,rounded
    corners,color=blue!50, inner
    xsep=2pt},background,label style={color=blue!50,label
    position=above,anchor=south,yshift=-0.2cm}]{{$n_B=2$}}
    \ket{0} &	&	\qw	        &	\gate[wires=2]{A(\theta_1,0)}	&	\qw	                                &	\qw	\\
    \ket{0} &	&	\gate{X}	&	\ghost{A(\theta_1,0)}	        &	\gate[wires=2]{A(\theta_2,0)}	&	\qw	\\
    \ket{0} &	&	\gate{X}	&	\qw	&	\ghost{A(\theta_3,0)}	       &	\qw	\\
    \ket{0} &	&	\qw	        &	 \qw       &	\qw & \qw & \\ \\
    \ket{0} &	&	\gate{X}	&	\qw	&	\qw	       &	\qw	\\
    \ket{0} &	&	\qw	        &	 \qw       &	\qw & \qw
\end{tikzcd}
    \caption{Simplified circuit obtained from the generic circuit shown in Fig. \ref{fig:gard} used to prepare a random set of initial states for neutrons for the case of 2 neutrons and 2 protons for $n_B=2$ (indicated in the blue area), and for the 3 neutrons and 3 protons on $n_B=3$. 
    The operator $A$ is used twice with only two random angles $\theta_1$ and $\theta_2$, while setting $\phi_i=0$. 
    A similar circuit is used to initialize the protons wave-function with two angles different from the ones used for neutrons. 
    We also show in this figure the circuit used to initialize the case of 3 neutrons and 3 protons on $n_B=3$ blocks (again, only qubits associated with neutrons are shown in the figure). }
    \label{fig:circ4}
\end{figure}

In the present work, we used a method that can introduce 
correlations between particles in the initial state, starting from a Slater determinant. Specifically, we adapt a method proposed in Ref. \cite{Gar20} to prepare a state preserving particle number. We show in Fig. \ref{fig:gard} a generic circuit acting on neutrons
that can give random unitary transformation of single-particle states preserving 
the number of neutrons, provided that we chose randomly a set of random angles 
$\{ \theta_\beta , \phi_\beta \}_{\beta=1, \Lambda}$, where $\Lambda$ can be increased at will. A similar set of operations with
a second set of angles can be used to randomize the initial states for protons. 
In practice, we realized that very few angles are needed to provide a sufficiently large number of initial states and use a simplified version of the circuit with only two applications of the $A$ operator, using two random angles $\theta_1$ and $\theta_2$ while setting $\phi_i=0$. The simplified circuit we finally used is shown in Fig. \ref{fig:circ4}. Note that these two angles don't correspond to entanglement between the initial neutrons and protons.

\begin{figure}[htbp]
  \raggedright
    \includegraphics*[width=0.43\textwidth]{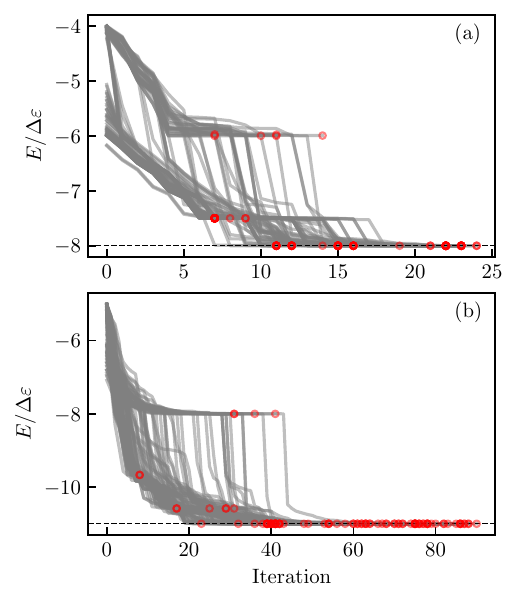}
\caption{Result obtained with the QEB-ADAPT-VQE starting from $100$ random initial conditions prepared using the circuit shown in Fig~\ref{fig:circ4} for (a) (2,2), and (b) (3,3). Different gray solid lines correspond to different trajectories in energy
as a function of the number of iterations. 
Here, the red open circles denote the position of the final iteration corresponding to each randomly chosen input state.}
  \label{fig:random}
\end{figure}

We show in Fig. \ref{fig:random} the result of $100$ events obtained using 
the random procedure described above to prepare a set of states. The QEB-ADAPT-VQE algorithm
is then applied for each of the initial states. The set of energy evolutions are systematically reported in Fig.  \ref{fig:random} for 2 protons and 2 neutrons in $n_B=2$ degenerated blocks (panel a), and 3 protons and 3 neutrons in $n_B=3$ degenerated blocks (panel b). We clearly see that the trajectories can be classified into groups of asymptotic energies (shown with open red circles). In panels a and b, 3 and 4 groups of final energies can be identified respectively. Noteworthily, the specific final energies $-7.5 \Delta \varepsilon$ and $-9.7 \Delta \varepsilon$, observed in panels a and b respectively, are the ones obtained in Fig. \ref{fig:adapt-4-on-2}-g and \ref{fig:adapt-6-on-3}-g. This 
indicates that the two initial states used previously belonged to the group of initial states that do not converge in the QEB-ADAPT-VQE towards absolute minimal energy. 
In the neutron-proton pairing case, we also see that many initial states finally lead to the proper convergence of the GS energy. For the events displayed in Fig. \ref{fig:random}, the success rate, i.e. the number of initial conditions converging to the lowest energy divided by the total number of states, is relatively high; it corresponds to $69\%$ and $73 \%$ respectively for panels a and b.

\begin{figure}[htbp]
    \centering
    \includegraphics[width=\linewidth]{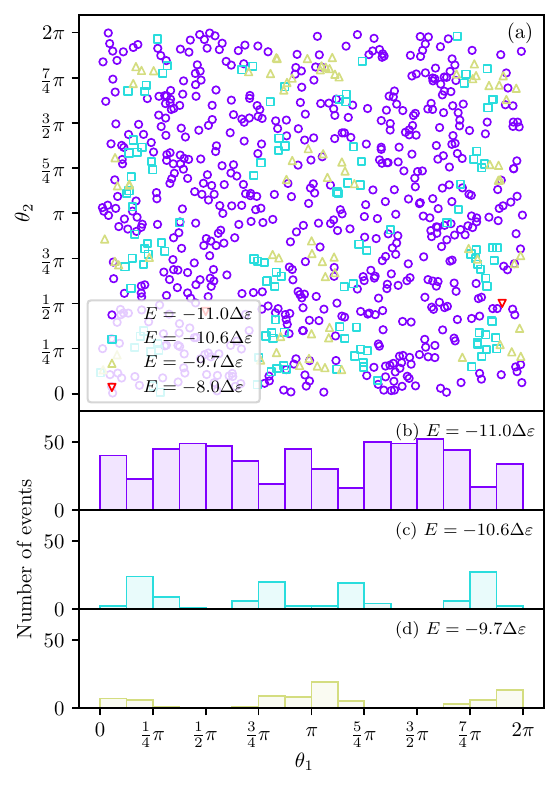}
    \caption{Panel a: Illustration of the different angles $(\theta^{(\lambda)}_1, \theta^{(\lambda)}_2)$ used to generate the  ensemble of initial states for the three blocks case using the circuit given in Fig. \ref{fig:circ4}. Here, $800$ events are shown. The ADAPT-VQE procedure is applied to each of these initial conditions. The different colors indicate which of the state presented in Fig. \ref{fig:random}-b, the ADAPT-VQE is converging to. The following convention is used for the final energy obtained: $E = -11.0 \Delta \varepsilon$ (purple circle), $E = -10.6 \Delta \varepsilon$ (blue square), $E = -9.7 \Delta \varepsilon$ (yellow triangle), and 
    $E = -8 \Delta \varepsilon$ (red triangle). In panels (b), (c) and (d) are shown a projection of the number of events on the $\theta_1$ axis respectively for the three lowest energies obtained. 
    Note that, the histogram for the final energy $-8 \Delta \varepsilon$ is not shown because it corresponds to too few events.  }
    \label{fig:thetascatterplot}
\end{figure}

We systematically studied the possible connection between the different subsets of initial states and the final asymptotic states obtained through 
the ADAPT-VQE procedure. Below, we summarize some conclusions drawn for the $n_B=3$ blocks case from this analysis. Similar conclusions apply to the $n_B=2$ case. First, regarding the different asymptotic states obtained in Fig. \ref{fig:random}: (i) The highest energy asymptotic state has energy $-8 \Delta \varepsilon$. This state corresponds to an excited state of the Hamiltonian. This asymptotic state is obtained only in rare cases where the initial state is strictly orthogonal to the ground state; (ii) The two asymptotic states at $-9.7 \Delta \varepsilon$ and $-10.6 \Delta \varepsilon$ are not eigenstates of the Hamiltonian. For these two states, we were not able to clearly identify a specific reason why the optimization procedure gets stuck in these energies; (iii) The asymptotic energy at $-11.0 \Delta \varepsilon$ corresponds to the proper ground state energy. It is worth mentioning that, although the proper energy is obtained, the final state of the ADAPT-VQE is not necessarily an eigenstate of $\langle {\bf S}^2 \rangle$ and $\langle {\bf T}^2 \rangle$ but mixes different channels. This stems from the ground state's degeneracy and the fact that the standard ADAPT-VQE technique relies only on a gradient criterion for the energy, without incorporating any criteria related to symmetries. Note that this is not necessarily a problem, since the mixing of different states can be removed by post-processing classical projection on the desired channels. Notably, in rare cases (less than $7 \%$ of the event shown in Fig. \ref{fig:random} that converge to the lowest energy), it might happen that the final proton or neutron numbers are $N=4$, $Z=2$ or $N=2$, $Z=4$. This is because the QEB pool ensures the total $A=N+Z$ particle conservation, but not that of $N$ and $Z$ individually. These states with different neutron and proton numbers are also degenerated with the ground state having $N=Z=3$ due to the strict spin-isospin degeneracy of single-particle states. Again, the rare events where this occurs can be easily identified simply by measuring the neutron or proton numbers after the ADAPT-VQE iterations. It is possible to introduce pools of operators that preserve the neutron and proton numbers separately, but we found that, by imposing too many conservation laws, the convergence properties could be slowed down and might degrade.

We also tried to identify specific patterns within subgroups of initial states 
that might explain the convergence towards specific asymptotic states but
did not find any of them. For instance, there is no evident correlations 
between the initial values of $\langle {\bf S}^2 \rangle$ and $\langle {\bf T}^2 \rangle$ and the final state. There is also no specific correlation between the convergence to the ground state and the overlap of the initial state with this GS, except for the case where the initial state is strictly orthogonal to the GS, which tends to converge to an excited state instead of the GS. The only clear tendencies we found are some correlations between the two angles $(\theta_1, \theta_2)$ used in the circuit given in Fig. \ref{fig:circ4} and the final state. This is illustrated in Fig. \ref{fig:thetascatterplot}, a scatter plot where we correlate the two angles' initial value, and the final energy obtained after optimization. In this figure, we clearly see that all regions in the $(\theta_1, \theta_2)$ plane are potentially able to converge toward the GS. The rate is enhanced around $\theta_1 = \pi/2$ and $3 \pi/2$, due to the absence of other competing final channels for these parameters. 

Besides this pattern seen in Fig. \ref{fig:thetascatterplot}, we see that the randomization procedure used here to build the initial state 
is an efficient way to avoid the arbitrariness associated with the initial state choice and has a high success probability of converging to the ground state. We conclude that this method can be considered a systematic and robust approach to obtain the GS energy for the neutron-proton pairing problem.

\section{Conclusion}

To prepare future applications in atomic nuclei, the neutron-proton pairing Hamiltonian problem is considered in the context of quantum computing. 
This research direction, by introducing both spin and isospin degrees of freedom on the same footing, constitutes a generalization of the problems involving only one species of particles, which have been recently investigated extensively using digital quantum computation. In the latter case, guided by classical techniques to treat superfluidity, the symmetry breaking/symmetry restoration strategy provided a rather efficient way to prepare quantum ansatzes on quantum computers. Before the study presented here, we put significant efforts into using a similar strategy for the neutron-pairing problem. Our first important conclusion is that the ansatz breaking the particle number symmetry with a set of other symmetries like total spins or isospin becomes extremely delicate to control in variational quantum 
algorithms. 

In the present work, as an alternative to the SR/SB approach, we focused on the ADAPT-VQE technique using different pools of operators. We used three different pools of operators and systematically investigated their ability to converge towards the ground state. In most situations, especially when spin-isospin blocks of singe-particle states are non-degenerated, we found that the ADAPT-VQE approach is powerful enough to properly solve the targeted ground state energy. This is a very encouraging result for future realistic shell model-like applications, where full degeneracy of neutron and proton single-particle states is very unlikely, at least due to the Coulomb interaction. A second important conclusion for future studies is that the number of iterations to achieve convergence might change significantly 
from one pool to another. Additionally, the symmetries that might or not be broken by a pool turns out 
to be an important ingredient for the convergence of the approach. 

In a few very specific cases where single-particle states are highly degenerated, we see that the ADAPT-VQE approach might lead to an energy differing from the GS energy, 
including some situations where the asymptotic ADAPT-VQE energy is lower than the exact GS energy. We identify that the latter case is due to the breaking of the particle number symmetry that might happen, in particular in the Qubit pool. We discuss two methods that control the broken symmetry during the energy descent.

Besides the symmetry-breaking problem, we introduce several techniques to improve the ADAPT-VQE approach when it was not converging in the first place. Among the techniques, we found that the embedding technique and the randomization of the initial state systematically solve the lack of convergence problems. We anticipate that these techniques will be useful for future applications in nuclear physics. 

\section{Acknowledgments }

DL thanks S. Baid and A. Ruiz Guzman for the discussions on the ADAPT-VQE at the early stage of this work. 
This project has received financial support from the CNRS through the AIQI-IN2P3 project and the P2IO LabEx (grant ANR-10-LABX-0038).  
JZ is funded by the joint doctoral programme of Universit\'e Paris-Saclay and the Chinese Scholarship Council.
This work is part of 
HQI initiative (\href{www.hqi.fr}{www.hqi.fr}) and is supported by France 2030 under the French 
National Research Agency award number ``ANR-22-PNQC-0002''.
We acknowledge the use of IBM Q cloud as well as the use of the Qiskit software package
\cite{Qis21} for performing the quantum simulations.

\end{document}